\documentclass[pra, superscriptaddress,showpacs,amsmath,amssymb,aps, twocolumn]{revtex4-1}

\usepackage[dvips]{graphicx}

\usepackage{color}
\usepackage{graphicx}
\usepackage{dcolumn}
\usepackage{bm}
\usepackage{xy}
\xyoption{matrix}
\xyoption{frame}
\xyoption{arrow}
\xyoption{arc}

\usepackage{ifpdf}
\ifpdf
\else
\PackageWarningNoLine{Qcircuit}{Qcircuit is loading in Postscript mode.  The Xy-pic options ps and dvips will be loaded.  If you wish to use other Postscript drivers for Xy-pic, you must modify the code in Qcircuit.tex}
\xyoption{ps}
\xyoption{dvips}
\fi

\entrymodifiers={!C\entrybox}

\newcommand{\bra}[1]{{\left\langle{#1}\right\vert}}
\newcommand{\ket}[1]{{\left\vert{#1}\right\rangle}}
\newcommand{\qw}[1][-1]{\ar @{-} [0,#1]}

\newcommand{\gate}[1]{*+<.6em>{#1} \POS ="i","i"+UR;"i"+UL **\dir{-};"i"+DL **\dir{-};"i"+DR **\dir{-};"i"+UR **\dir{-},"i" \qw}

\newcommand{\multigate}[2]{*+<1em,.9em>{\hphantom{#2}} \POS [0,0]="i",[0,0].[#1,0]="e",!C *{#2},"e"+UR;"e"+UL **\dir{-};"e"+DL **\dir{-};"e"+DR **\dir{-};"e"+UR **\dir{-},"i" \qw}
\newcommand{\ghost}[1]{*+<1em,.9em>{\hphantom{#1}} \qw}

\newcommand{\lstick}[1]{*!R!<.5em,0em>=<0em>{#1}}

\newcommand{\Qcircuit}{\xymatrix @*=<0em>}

\begin{document}

\title{Quantum autoencoders via quantum adders with genetic algorithms}

\author{L. Lamata}\email{lucas.lamata@gmail.com}
\affiliation{Department of Physical Chemistry, University of the Basque Country UPV/EHU, Apartado 644, 48080, Bilbao, Spain}
\author{U. Alvarez-Rodriguez}
\affiliation{Department of Physical Chemistry, University of the Basque Country UPV/EHU, Apartado 644, 48080, Bilbao, Spain}
\affiliation{Basque Centre for Climate Change (BC3), Sede Building, Campus EHU/UPV, Leioa, Bizkaia, Spain}
\affiliation{School of Mathematical Sciences, Queen Mary University of London, London E1 4NS, UK}
\author{J. D. Mart\'in-Guerrero}
\affiliation{IDAL, Electronic Engineering Department, University of Valencia, Avgda. Universitat s/n, 46100 Burjassot, Valencia, Spain}
\author{M. Sanz}
\affiliation{Department of Physical Chemistry, University of the Basque Country UPV/EHU, Apartado 644, 48080, Bilbao, Spain}
\author{E. Solano}
\affiliation{Department of Physical Chemistry, University of the Basque Country UPV/EHU, Apartado 644, 48080, Bilbao, Spain}
\affiliation{IKERBASQUE, Basque Foundation for Science, Maria Diaz de Haro 3, 48011, Bilbao, Spain}
\affiliation{Department of Physics, Shanghai University, 200444 Shanghai, China}

\begin{abstract} 
The quantum autoencoder is a recent paradigm in the field of quantum machine learning, which may enable an enhanced use of resources in quantum technologies. To this end, quantum neural networks with less nodes in the inner than in the outer layers were considered. Here, we propose a useful connection between quantum autoencoders and quantum adders, which approximately add two unknown quantum states supported in different quantum systems. Specifically, this link allows us to employ optimized approximate quantum adders, obtained with genetic algorithms, for the implementation of quantum autoencoders for a variety of initial states. Furthermore, we can also directly optimize the quantum autoencoders via genetic algorithms. Our approach opens a different path for the design of quantum autoencoders in controllable quantum platforms.
\end{abstract}

\maketitle

\section{Introduction} 
Quantum machine learning is an emerging field that aims at enhancing machine learning methods with quantum technologies~\cite{Schuld,Biamonte,Dunjko,DeWolf}.  The synergy works two-fold: either through genuine quantum effects, such as entanglement, to speed up the calculations of machine learning~\cite{Schuld,Biamonte}, or to employ classical machine learning to improve quantum processes~\cite{ga7,Sciarrino1}. In this respect, an advanced protocol has been considered, inspired in the classical autoencoder techniques of deep learning~\cite{Goodfellow}, namely, a quantum autoencoder~\cite{Kim,Aspuru}. Other topics related to biomimetic quantum technologies in general, which have emerged in recent years, involve quantum artificial life~\cite{Unai1,Unai2}, quantum reinforcement learning in quantum technologies~\cite{Lamata2017}, quantum memristors~\cite{Sanz,Mikel2,Mikel3,NoriMemristors}, quantum Helmholtz and Boltzmann machines~\cite{Perdomo1,Perdomo2,Perdomo3}, and quantum machine learning with time-delay equations~\cite{UnaiTimeDelay1,UnaiTimeDelay2}.

\begin{figure}
{\includegraphics[width=0.5\textwidth]{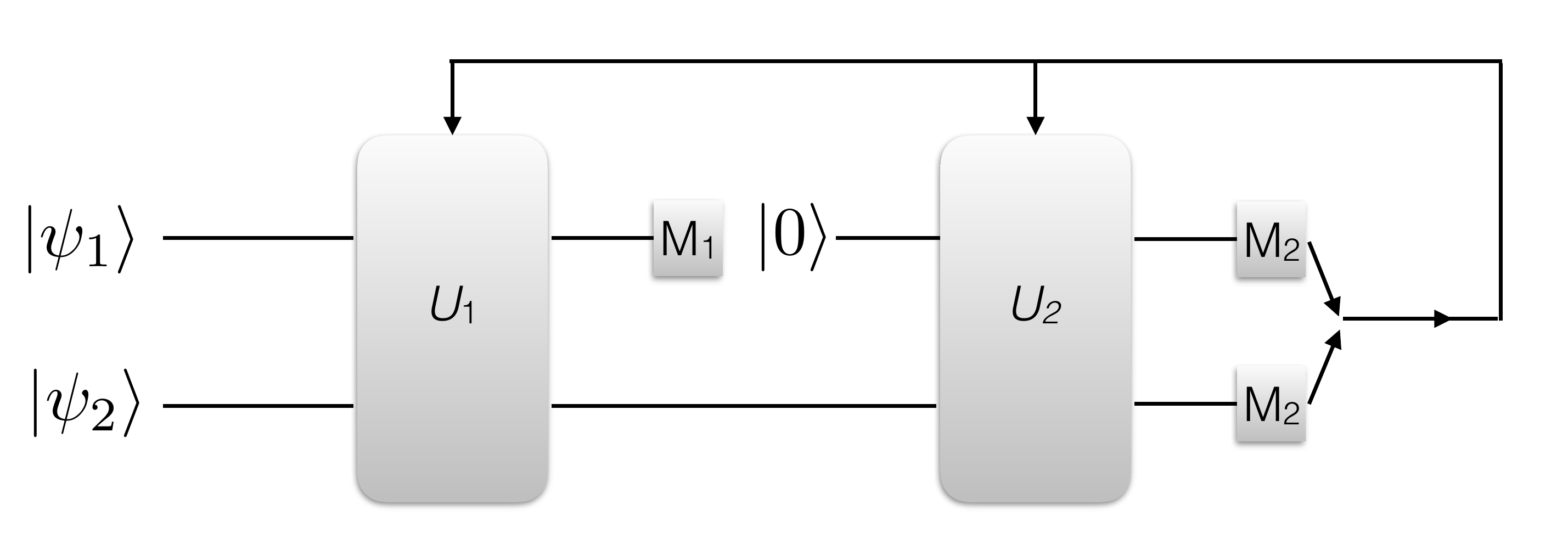}}
\caption{ Scheme of a 2-qubit quantum autoencoder as proposed in Refs.~\cite{Kim,Aspuru}. It is based on a quantum neural network where the outer layers (two-qubit input and output states) contain one more qubit than the inner layer (single-qubit state in between the unitary gates $U_1$ and $U_2$) and the former are connected to the latter via the encoder $U_1$ and the decoder $U_2$. These operations are optimized, via, e.g., gradient descent techniques, with controllable parameters by means of a classical feedback loop, in order that the output state maximizes the overlap with the input state. $M_1$ is a dummy measurement, while $M_2$ is a computational basis measurement on the output two-qubit state that is employed to close the feedback loop. After convergence, $U_1$ is the optimized encoding operation.}\label{QAutoencoder}
\end{figure}

A quantum autoencoder, see Fig.~\ref{QAutoencoder}, is a quantum device which can reorganize the quantum information of a subset of a Hilbert space spanned by a basis of initial $n$-qubit states onto a subset of a Hilbert space spanned by $n'<n$ qubit states. Therefore, in this way, one may encode the information into a smaller amount of resources, which may be useful in different quantum technologies. Quantum autoencoders, at variance with previous results on compression of quantum information~\cite{Sch95,NielsenChuang}, aim at reshuffling the information to employ less resources, similarly to defragmentation in classical computing. To achieve this reordering, one may employ a feedforward quantum neural network~\cite{Kim}, which requires a smaller amount of qubits in the inner layer than in the input and output layers. Via gradient descent techniques, we can optimize the intermediate transition unitary operations between layers such that, for each input state, the output state has almost perfect overlap with it. Therefore, the first part of the network (encoder), which reorganizes the input states onto the inner layer of qubits, contains the desired solution to the problem due to the discarding of superfluous information. Classical autoencoders usually employ that useful information to make feasible the training of deep neural networks and to extract high-level features that describe the original data space correctly using a lower dimension.

In the last years, the design and implementation of approximate quantum adders~\cite{qa1,qa2,qa3,qa3bis,qa3bis2,qa4,qa5,qa4bis,qa4bis2} has emerged with unexpected connections. Recently, it was proven that a quantum operation that adds two unknown quantum states is forbidden in general~\cite{qa1,qa2}. Since then, important results in optimized approximate or probabilistic quantum adders have been analyzed, and in some cases implemented, which may find applications to different aspects of quantum technologies~\cite{qa1,qa2,qa3,qa3bis,qa3bis2,qa4,qa5,qa4bis,qa4bis2}. 

\begin{figure}
{\includegraphics[width=0.5\textwidth]{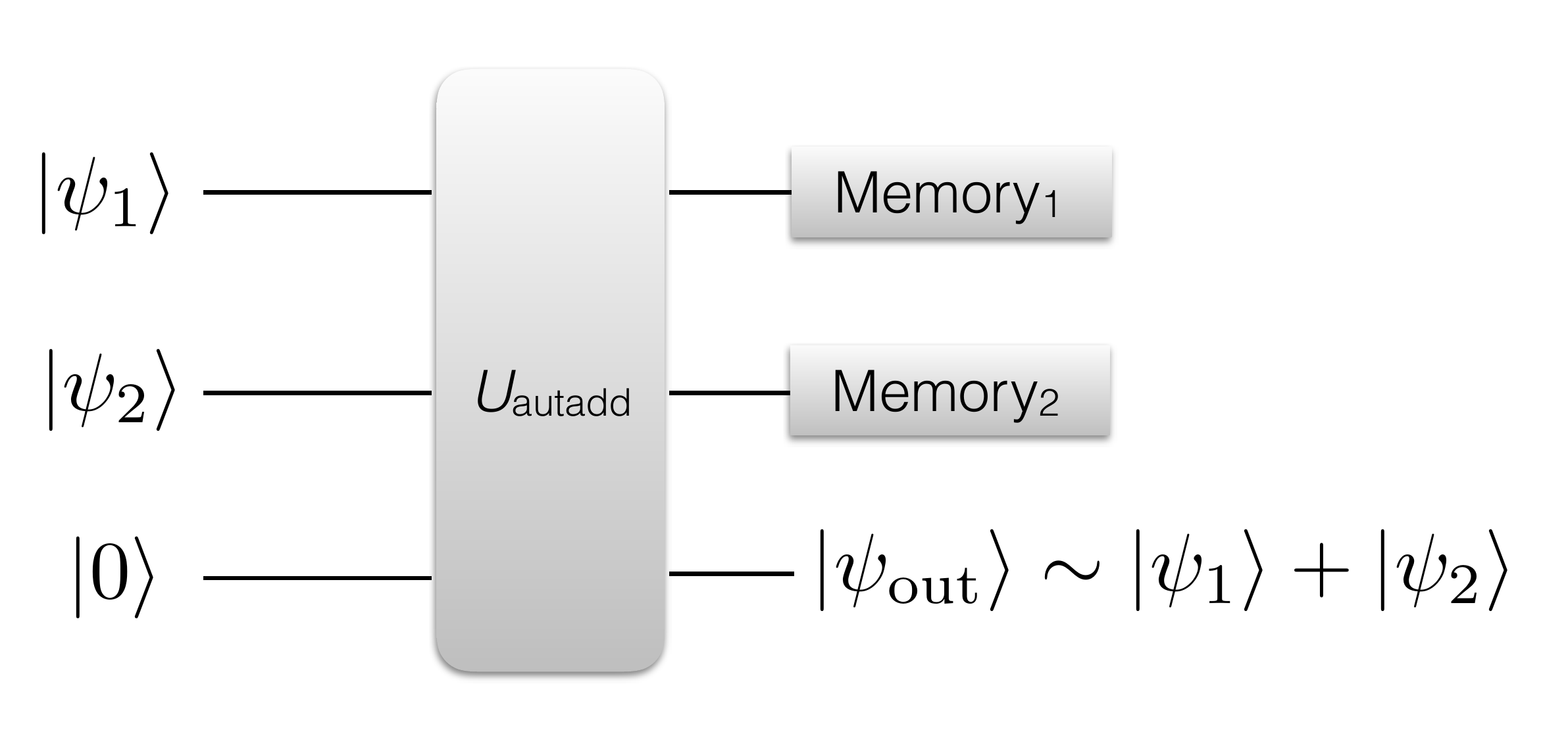}}
\caption{ Scheme of our proposed 2-qubit quantum autoencoder. It is based on a 3-qubit device composed of an optimized approximate quantum adder operation $U_{\rm autadd}$ that acts on the input states, $|\psi_1\rangle$ and $|\psi_2\rangle$, as well as an ancilla qubit $|0\rangle$. The approximate quantum adding operation produces a state with large overlap to $|\psi_1\rangle+|\psi_2\rangle$, allowing for an encoding of the initial 2-qubit quantum information subspace into a single-qubit subspace, to a certain fidelity. The memory units allow for maintaining entanglement of the three output qubits and, after possible local operations on $|\psi_{\rm out}\rangle$, under proper circumstances, to invert the encoding via $U^\dag_{\rm autadd}$.}\label{QAutoencoderAdder}
\end{figure}

In this article, we propose and analyze a connection between a quantum autoencoder and an approximate quantum adder, see Fig.~\ref{QAutoencoderAdder}. The main insight in this respect is the realization that, provided a perfect quantum adder, a perfect quantum autoencoder and perfect quantum compressor would be attained. The reason for this is that, given a hypothetical unitary operation $U$ able to add two unknown quantum states, $U|\psi_1\rangle|\psi_2\rangle\propto|\psi_1\rangle+|\psi_2\rangle$, a trivial encoding would be produced. In this sense, $U$ would allow us to map the quantum information of tensor products of two, or by iteration of $n$, single-qubit states onto a single qubit. Then, in order to decode the state, one would apply the inverse of $U$, i.e., $U^\dag$, retrieving the initial state. By linearity, this would also apply to superpositions of the initial entries. We point out that this would naturally imply that arbitrary highly entangled states could be mapped onto a single qubit state, which in general is, of course, unfeasible. Nevertheless, via approximate quantum adders one may encode entangled states onto lower dimension states, while, by employing entangled ancillas, one may reverse the protocol, performing the decoding, and retrieve the initial entangled state with fidelity one. As proved in Ref.~\cite{qa1}, an ideal quantum adder is forbidden by the laws of quantum mechanics. Another complementary proof is given by this connection, because if an ideal and universal $n$-qubit quantum adder existed, reorganization, as well as compression of quantum information from $n$-qubit states to a single one would be feasible, which is clearly forbidden in general because of different Hilbert-space dimensionality and because it would violate Schumacher's theorem~\cite{Sch95}. Despite this, one can employ the insight obtained on approximate quantum adders optimized with genetic algorithms~\cite{qa5} to propose quantum autoencoders that rely on the previous optimization. A further possibility is to optimize the protocol of quantum autoencoders via genetic algorithms, using similar techniques as in previous works~\cite{ga7,qa5}. In this article, we will follow both approaches. After revising the concept of approximate quantum adders in Section \ref{ApproxQAdder}, we consider the possible implementation of quantum autoencoders via approximate quantum adders in Section \ref{QAutoApQA}. Later on, we study quantum autoencoders optimized with genetic algorithms that employ a restricted amount of gates and in this context achieve also high fidelities, in Section \ref{QAutoGA}. We then analyze proposals for quantum implementations with trapped ions, superconducting circuits, and quantum photonics, in Section \ref{ProposalsImplem}. Finally, we give our Conclusions in Section \ref{Conclus}.

\section{Approximate Quantum Adders}\label{ApproxQAdder}

A decomposition of approximate quantum adders in terms of single and two-qubit gates can be obtained according to diverse criteria. Some adders may be obtained by defining them to add optimally the computational basis states, and extended by linearity to superpositions of these. Other quantum adders can be optimized by genetic algorithms in the context of limited available amount of gates. By employing the proposed protocol of Fig.~\ref{QAutoencoderAdder}, we describe now how to implement a quantum autoencoder based on each of these cases. For simplicity, the unitary gate denoted as $U_{\rm autadd}$ in the figure will be called $U$ in the following example.
 
\subsection{The basis quantum adder}

This quantum adder was defined~\cite{qa1,qa5} to perfectly add the basis states, and approximately the superposition states, according to

\begin{align}\nonumber
&U \ket{000}=\ket{000}, \hspace{0.25cm} U \ket{010}=\ket{01+}, \hspace{0.25cm} U \ket{100}=\ket{10+},\\\nonumber & U \ket{110}=\ket{001}, \hspace{0.25cm} U \ket{001}=\ket{110}, \hspace{0.33cm} U \ket{011}=\ket{01-},\\ & U \ket{101}=\ket{10-}, \hspace{0.15cm} U \ket{111}=\ket{111},\label{basisqaddergates}
\end{align}
where the first two qubits in the lhs are the addend states, the third one is an ancillary qubit, and the last qubit in the rhs is the outcome state of the addition. In the protocol the initial ancilla is always the $\ket{0}$ state, but the previous definition of the $U$ quantum adder includes all possible states for univocally defining the gate.

The fidelity of the quantum adder $U$ is defined in terms of the output state $\rho_{\rm out}$ as
\begin{eqnarray}
&& F={\rm Tr}(\ket{\Psi_{\rm id}}\bra{\Psi_{\rm id}}\,\rho_{\rm out}) \label{fid}, \\ \nonumber && \rho_{\rm out}={\rm Tr}_{12}(U\ket{\psi_1}\bra{\psi_1}\otimes\ket{\psi_2}\bra{\psi_2}\otimes\ket{0}\bra{0}U^\dag)
\end{eqnarray}
where $\ket{\Psi_{\rm id}}\propto \ket{\psi_1}+\ket{\psi_2}$ is the ideal outcome of the sum, and the trace is taken over the first two qubits. In Ref.~\cite{qa5}, a subset of all possible input states were considered in order to make the approximate quantum adder consistent, namely, $|\psi_i\rangle$, $i=1,2$ were defined as
\begin{align}
\ket{\psi_i} &= \left(\begin{array}{c}
\cos\, \theta_i \\
\sin\,\theta_i 
\end{array}\right),\label{defIntervalPsi}
\end{align} 
and the angles $\theta_i\in\{0,\pi/2\}$.

The average fidelity of this region is $94.9\%$, while the lowest fidelity is $85.4\%$.

\section{Quantum autoencoder based on approximate quantum adders}\label{QAutoApQA}

 We propose a quantum autoencoder based on the previous approximate quantum adder following the scheme of Fig.~\ref{QAutoencoderAdder}. For any input, two-qubit product state in the considered region defined by Eq.~(\ref{defIntervalPsi}), i) apply the quantum adder given by $U$ onto this state and an ancilla $\ket{0}$ state, ii) store the first two outputs onto quantum memories, and employ the approximate addition of the input states onto $\ket{\psi_{\rm out}}$ for any desired quantum task, including quantum communication and single-qubit gates, iii) retrieve the decoded, modified, two-qubit state via application of $U^\dag$ onto the memory qubits and the output state. By linearity, the protocol can be extended to superpositions of input states. A criterion that should hold for consistency is that the two memory qubits should be in the same quantum state for the considered input states after the quantum adding operation. This way, the effect of any local operation on the encoded $\ket{\psi_{\rm out}}_i$ that acts differently on the latter for different $i$, where $i$ refers to the specific input $\ket{\psi_1}_i\ket{\psi_2}_i$ state considered, can be efficiently retrieved onto the corresponding two-qubit gate on $\{\ket{\psi_1}_i\ket{\psi_2}_i\}$ via $U^\dag$.  For cases in which, for different $i$, the memory qubit states differ, the encoding operation will work up to a certain fidelity, which will decrease with the distance between these memory states and also depend on the specific adding operation and input states.

We point out that in the case of a quantum autoencoder based on an approximate quantum adder, the usage of quantum memories for the output, first two entries are necessary. The reason is that the inverse operation of the adder, $U^\dag$, may retrieve the input two-qubit subspace with higher fidelity, as far as the $\ket{\psi_{\rm out}}$ state is controllably modified in the process. In order to map an original two-qubit gate in the initial subspace onto the single-qubit one, one has to encode it according to the specific quantum adder employed. In the case here proposed, for example, if one wants to apply a CPHASE gate onto the inputs $\ket{00}$ and $\ket{11}$, one would employ the mapping of Eq.~(\ref{basisqaddergates}). Realizing that in this case both memory qubits would take the same values for both input states, i.e., $\ket{00}$, one would just need to apply a phase gate on $\ket{\psi_{\rm out}}$ to introduce a minus sign between $\ket{0}$ and $\ket{1}$ single-qubit states. Later on, one would apply the inverse adding operation, $U^\dag$, retrieving the input states with a minus sign on the $\ket{11}$ state, as corresponds to the correct application of the CPHASE gate.

In general, the encoding of a desired two-qubit gate operation onto a single qubit gate in the compressed space will depend both on the gate and on the subspace on which it acts. Clearly, a two-qubit gate acting on the complete two-qubit Hilbert space cannot be mapped onto a single-qubit gate without losing information. But the previous example can be useful to illustrate that, in some cases, the complexity of a two-qubit gate implementation may be reduced to a single-qubit gate. This may be useful, for example, if we have full availability of a universal quantum computer in some lab, where we can implement the quantum adder and store the memory qubits, but only availability of single-qubit gates in some other lab, to which we send the $\ket{\psi_{\rm out}}$ qubit via a quantum communication channel. Another possibility where this approach can be useful is a situation in which one considers quantum information encoded in qudits but has only access to single-qubit gates. The generalization of the basis adder to qudits, with a lower bound on the minimal dimension of the ancillary state, was presented in Ref.~\cite{qa5}. In these cases, the encoding via approximate quantum adders may also prove fruitful. This formalism could also be employed as a useful transducer between a processing and a storage unit.

With the previous considerations in mind it would be useful to study a generalization of the quantum adder to extend the protocol presented here. An important limitation of quantum adders, if employed as autoencoders, is the reduced amount of inputs they can handle. In order to avoid this restriction we propose a quantum multi-adder, $U_n$, with the capacity of achieving the normalized sum, $|\psi_s\rangle$, of $n$ input quantum states given as a tensor product. Notice that the general multi-adder is forbidden by the simple fact that it has the original quantum adder as a particular case.
\begin{equation}
U_n \left( \bigotimes^{n}_i |\psi_i\rangle \right) |A\rangle = |\psi_s\rangle |B\rangle
\end{equation}

The compression of information would be enhanced by an ideal multi-adder $U_n$, as it would enable to encode the information of any number of quantum states into a single one. Even if the ideal operation is forbidden, an analogous analysis to the one that has been carried out in the previous section would clarify the potential of approximate multi-adders to reduce resources in quantum information processing tasks. 

We can already speculate about the selection of a particular multi-adder according to the properties of the protocol to be implemented. Firstly, we have to be aware of the number of degrees of freedom, as the number of independent parameters, for a selection of the dimension of Hilbert space in which the quantum states, $|\psi_i\rangle$, are described. And secondly, we should estimate the part of information that is relevant for the problem we are dealing with. Therefore, the adder to use will depend on the type and amount of information that the second layer of states needs. The specific operations which may be efficiently encoded for a certain quantum adder will possibly depend on symmetries of these operations, and of the states for which the quantum adder is optimized. For example, the basis adder presented above can be useful to compress the information of the observable $\sigma_z$, with an ideal fidelity, as we show in the following calculation:

Let $|\psi_i\rangle = \cos \theta_i |0\rangle + \sin \theta_i |1\rangle$, $i \in \{ 1,2\}$, be any pair of states fulfilling the conditions to be processed by the basis adder. The $\sigma_z$ statistics of the product of both states, $\langle \sigma_z \rangle_{\rho_1 \rho_2}$, with $\rho_i = |\psi_i \rangle\langle \psi_i|$, is recovered from $|\psi_s\rangle$ the summand of the operation. This encoding is highly nontrivial, given that it enables the measurement of a nonlinear operation. We first apply the basis adder to the pair of states we have selected,
\begin{eqnarray}
&&U |\psi_1\rangle |\psi_2\rangle |0\rangle = \cos\theta_1 \cos\theta_2 |000\rangle + \cos\theta_1 \sin\theta_2 |01+\rangle\nonumber\\ &&+ \sin\theta_1\cos\theta_2 |10+\rangle + \sin\theta_1\sin\theta_2 |001\rangle,
\end{eqnarray}
and trace out the input qubits to analyze the summand $\rho_s$
\begin{eqnarray}
\rho_s = && |0\rangle\langle 0| ( \cos^2 \theta_1 \cos^2 \theta_2 )  + |1\rangle\langle 1| ( \sin^2 \theta_1 \sin^2 \theta_2 ) + \nonumber\\ && |+\rangle\langle +|(\cos^2\theta_1 \sin^2\theta_2 + \sin^2\theta_1 \cos^2\theta_2)\nonumber\\&& + (|0\rangle\langle 1| + |1\rangle\langle0|)\cos\theta_1\sin\theta_1\cos\theta_2\sin\theta_2 .
\end{eqnarray} 
One can now understand the nonlinear encoding the adder implements in terms of the observables and states involved in the process. 
\begin{equation}
\langle \sigma_z \rangle_{\rho_s} = \cos^2 \theta_1 \cos^2 \theta_2 - \sin^2 \theta_1 \sin^2 \theta_2 = \langle \sigma_z \rangle_{\rho_1 \rho_2}
\end{equation}

Summarizing, we believe that a further exploration will show a correlation between the type of adder and the combination of observables whose information is perfectly encoded in the reduced expression. However, this analysis goes beyond the scope of this manuscript.

\section{Quantum autoencoders via genetic algorithms}\label{QAutoGA}
As pointed out above, approximate quantum adders can also be optimized with genetic algorithms. Following the previous approach for the basis adder, these quantum adders can also be employed to define quantum autoencoders. The formalism of genetic algorithms allows one to define the maximum number of gates in the desired decomposition, as well as its structure of single- and two-qubit gates~\cite{ga7,qa5}. Therefore, it can be useful for practical purposes in current or near future quantum implementations with trapped ions, superconducting circuits, and quantum photonics, which employ a limited amount of resources. In order to define a quantum autoencoder via an optimized quantum adder, we propose to follow the scheme on Fig.~\ref{QAutoencoderAdder}, where now the unitary gate $U_{\rm autadd}$ will be given by the corresponding quantum adding operation. 

An alternative approach is to directly optimize the quantum autoencoders with genetic algorithms following the scheme of Fig.~\ref{QAutoencoder}, instead of previously employed optimization methods, e.g., gradient descent. An advantage of genetic algorithms is that local minima can in principle be better avoided in some situations. The first step is to design a function that relates the ``genetic code'' of each ``individual'' in the program to a specific autoencoder, either $U_{\rm autadd}$ or the pair of $U_1$ and $U_2$ in the more general case. The genetic code is represented by a $g \times 3$ matrix where $g$ is the number of gates. Accordingly, each row provides specific information about the gate to implement. The interaction type and phase are encoded in the first two columns, while the third one determines the qubit to act upon. The universal set of gates we are working with is given by single-qubit rotations and the M{\o}lmer-S{\o}rensen gate: $\{ R_{x}(\theta), R_{y}(\theta), R_{z}(\theta), e^{-i \sigma_y \otimes \sigma_y \theta/2}  \}$. In each loop a new generation is bred, where the individuals are hierarchically recombined, mutated and selected according to the fidelity of the corresponding autoencoder. We have followed this approach for a certain set of cases via numerical simulations, and obtained consistent results. Namely, when compressing a set of two-qubit states which contains at most two linearly-independent states, the compression is optimal and the protocol converges to fidelity one onto a single-qubit subspace. When trying to compress four orthogonal two-qubit states onto two single-qubit ones, a $50\%$ average fidelity is obtained. For three orthogonal states, an average fidelity of 2/3 is obtained.  For non orthogonal four-qubit or three-qubit states, the fidelity will be larger than for the corresponding orthogonal ones, the difference depending on the overlap with each other. More precisely, in order to test the validity of our algorithm we have analyzed an optimized autoencoder for a nontrivial case involving a set of three states: $|\psi\rangle_1 = \cos \frac{\pi}{3} |00\rangle + \sin \frac{\pi}{3} |01\rangle$, $|\psi\rangle_2 = |01\rangle$, $|\psi \rangle_3= \cos \frac{\pi}{8} |10\rangle -i \sin \frac{\pi}{8} |11\rangle$. We have achieved an average fidelity of $87.51 \%$ in the general case of $U_1 \neq U_2^{\dag}$ for a reduced number of gates. In Table \ref{ga} we show the result of the Genetic Algorithm. This is afterwards simplified to reduce the number of gates following the procedure of combining the phases of two consecutive gates of the same type acting on the same qubit. See the following quantum circuit diagrams for the specific decomposition of the autoencoder in quantum gates. We point out that an interesting feature of genetic algorithms is precisely that they already provide the sequence of elementary gates generating the solution.

\begin{align}\label{qcd1}
    \Qcircuit @C=1em @R=.7em {
\lstick{\ket{\psi_1}} & \gate{R_x} & \gate{R_z} & \multigate{1}{MS} & \qw & \multigate{1}{MS} \\
\lstick{\ket{\psi_2}} & \gate{R_x} & \qw & \ghost{MS} & \gate{R_z} & \ghost{MS} 
}  
 \end{align}

\begin{align}\label{qcd2}
    \Qcircuit @C=1em @R=.7em {
\lstick{\ket{\psi_1}}  & \qw & \multigate{1}{MS} & \gate{R_y}  & \qw & \qw  \\
\lstick{\ket{\psi_2}} & \gate{R_x}  & \ghost{MS} & \gate{R_z} & \gate{R_y} & \gate{R_x} 
}  
 \end{align}

Here, $R_{x,y,z}$ represent the single qubit rotations, while $MS$ is the M{\o}lmer-S{\o}rensen gate. The diagrams show the type of obtained interactions, not specifying the phases, for $U_1$ and $U_2$ respectively. The algorithm we employed is a variant of a previous program \cite{ga6,ga7,qa5}. 

\begin{table}[h!] 
\centering
\begin{tabular}{*{9}{|c}|}
\hline
Gate  & 1 & 3 & 1 & 1 & 3 & 4 & 3 & 4 \\
\hline
Phase & 1.575 & 2.176 & 0.246 & 0.493 & 4.099 & 5.214 & 2.719 & 5.594 \\
\hline
Qubit & 1 & 1 & 2 & 2 & 1 & - & 2 & - \\
\hline
\hline
Gate & 1 & 4 & 2 & 3 & 2 & 2 & 2 & 1 \\
\hline
Phase & 2.672 & 5.569 & 1.976 & 3.783 & 0.963 & 2.823 & 2.930 & 1.315 \\
\hline
Qubit & 2 & - & 1 & 2 & 2 & 1 & 1 & 2 \\
\hline
\end{tabular}
\caption{{\bf Result of the Genetic Algorithm.} The first three rows correspond to $U_1$ and the last three to $U_2$. The gates $\{1,2,3,4\}$ correspond to $\{\sigma_x, \sigma_y , \sigma_z, MS\}$, the phases are presented in radians and rounded to three decimal digits. }
\label{ga}
\end{table}

\section{Proposals for quantum implementations}\label{ProposalsImplem}

In the following, we describe possible implementations with trapped ions, superconducting circuits, and quantum photonics.

\subsection{Trapped Ions} In trapped-ion quantum platforms, linear chains of ions are confined via electromagnetic fields~\cite{ion}. Both internal electronic states, and quantized motional degrees of freedom, are available for quantum control. The Hamiltonian describing the coupling between an ion and a laser, involving internal states and a motional mode, in the Lamb-Dicke regime, is given by
\begin{equation}
\label{ion}
H= \hbar \Omega \sigma^{+} \left[ 1 + i \eta \left( a e^{-i \nu t}+ a^{\dag} e^{i \nu t} \right) \right]e^{i(\phi-\delta t)} + \textrm{H.c.}
\end{equation}
Here,  $\sigma^{+}$ is the spin raising operator, $\eta$ the Lamb-Dicke parameter, $\Omega$ is the Rabi frequency, $a$ and $a^{\dag}$ the motional annihilation and creation operators, $\nu$ the trap frequency,  $\phi$ the laser field phase, and $\delta$ the detuning between the laser and qubit frequencies.

In our proposal, each two-qubit state can be encoded either in four metastable levels of a single ion or two levels of a pair of ions, and an additional ion serves as an ancillary qubit in the case of autoencoder with quantum adder. The necessary universal gate set, $\{\sigma_x, \sigma_y , \sigma_z, MS\}$, can be implemented with available M\o lmer-S\o rensen gates and sequences of single-qubit gates \cite{moso} in trapped ions, allowing us to implement our approximate quantum adders and therefore the quantum autoencoders. The memory qubits can be stored in long-lived internal electronic states.

\subsection{Superconducting Circuits} Superconducting circuits are made of superconducting elements as microwave cavities, as well as nonlinear elements, i.e., Josephson junctions, which allow one to design effective quantum two-level systems~\cite{scc}. Via standard circuit quantization, one can often obtain a Hamiltonian resembling the Jaynes-Cummings model, 
\begin{equation}
\label{scc}
H= \omega a^{\dag} a + \frac{\omega_0}{2} \sigma^z + g(a \sigma^{+} + a^{\dag} \sigma^{-}),
\end{equation}
where $\omega$ is the microwave mode frequency, $g$ is the photon-qubit coupling constant, and $\omega_0$ is the qubit frequency, which is encoded in the quantum excitations of the effective two-level system. 

Transmon qubits have long coherence times, and therefore they can be especially suitable for this kind of encoding. A universal gate set including entangling gates can be carried out with fidelities around 99\%, e.g., via capacitive coupling, or via resonators, as in Eq.~(\ref{scc}) when considering more than one qubit. In Ref.~\cite{qa5}, an optimized approximate quantum adder obtained via genetic algorithms was carried out experimentally in the IBM Quantum Experience processor, obtaining a good agreement with theoretical simulations. Therefore, it is expected that quantum autoencoders based on approximate quantum adders, or on direct optimization via genetic algorithms, may be achieved with current technology.

\subsection{Quantum Photonics} Quantum photonic platforms are appropriate for implementing quantum autoencoders via quantum adders, given that probabilistic quantum adders have already been performed in some photonic laboratories, e.g., Ref.~\cite{qa3}. In this quantum platform, qubits may be encoded in polarization or dual rail states,  and a universal gate set performed via linear optics elements in combination with ancillary qubits and detectors~\cite{pho}. The fact that photons are the best quantum information carriers for long quantum communication, makes them interesting for implementing quantum autoencoders via quantum adders and genetic algorithms, given that having to transmit a smaller amount of quantum information through a quantum channel can save resources.

\section{Conclusions}\label{Conclus}

 We have proposed and analyzed a connection between approximate quantum adders and quantum autoencoders, which may be useful to reorganize or compress quantum information employing less resources. This connection allows one to implement optimized approximate quantum adders with genetic algorithms that are mapped onto quantum autoencoders, as well as a direct optimization of quantum autoencoders with genetic algorithms. We point out that, as compared to previous proposals for quantum autoencoders~\cite{Kim,Aspuru}, our protocol can be performed in a single shot, and does not require training. Namely, once one has optimized the corresponding quantum adder, it can be employed for a quantum autoencoder from the very beginning, without the need to learn the structure of the input states as in previous works. The tradeoff is that the fidelity will not be 1 in general, and will depend, for each quantum adder, on the specific states considered. The employment of quantum memories in our protocol is also not necessary if one is only concerned with the compression (encoding) part, and not on the decompression (decoding). Indeed, in standard quantum autoencoders this is always the case, namely, one only wants to compress the information and not to go back onto the original space. In this scenario, we do not need to consider the quantum memories, and just employ the smaller amount of registers with the adding operation, which will work to an optimized fidelity. Nevertheless, being our protocol unitary with the presence of the memories, this will allow, if desired, to retrieve the original subspace. 
 
The quantum adder operation, being unitary, can in general be applied to any input quantum states. Of course, depending on the quantum adder optimization, the fidelities for adding some quantum states will be larger, while for other quantum states, will be lower. An efficient way of deploying a quantum autoencoder via quantum adders would be to employ a quantum adder that is known to add better the family of input states that one would like to compress. Even in the case for which this full information is not available a priori, if partial information of this family exists then the quantum adder can still be optimized under these constraints.
 
Regarding a comparison of resources between quantum autoncoders based on quantum adders and those based directly on classical optimization, an exact description cannot be provided given that it will strongly depend on the specific cases involved, namely, the input state set of the classically optimized autoencoder, including Hilbert space dimensionality, as well as unitary set available, in previously developed quantum autoencoder paradigms. In the case of the quantum autoencoder based on quantum adders, the performance will rely on the resources needed for the quantum adder, which also include the set of states for which the adding operation is of maximal fidelity, either the employment of memories or not, depending on whether decoding is required, and the like. Therefore, a fully quantitative comparison is not feasible, given that it will change on a case-by-case basis. However, it is natural to expect that there will be instances where the quantum autoencoder based on quantum adder will perform better than the one based on classical optimization, and cases in which it will be worse. The concept of the quantum autoencoder based on quantum adder is basically a probabilistic, single-shot quantum autoencoder whose performance depends on the fidelity of the underlying adding operation for specific quantum states.
 
We point out that the number of states which can be perfectly encoded will in general depend on the dimensionality of the encoding system: for two input states, a single qubit will suffice, given that it has dimension two. In this case, the memory qubits can also be independent of the regarded input states (although they will in general depend on the remaining states of the input complete Hilbert space). This is deduced from an isomorphism argument, namely, a unitary operation maps an orthonormal set onto an orthonormal set. Thus, two input orthonormal states can be mapped via a unitary operation onto two encoded orthonormal states. For, say, three  or four 5-qubit input states, a 2-qubit compressed space is enough for perfect encoding, given that it has dimensionality four. Moreover, the remaining three memory qubits can be independent of the four input states, given that a similar isomorphism argument as before guarantees that four input orthonormal states can be mapped onto four encoded orthonormal states. For non-orthogonal input states, or for other states in the input complete Hilbert space, the fidelity will in general be lower, and depending on the specific states under consideration. This dimensionality argument refers to the employment of memories for decoding after the protocol, but in general the final fidelities will strongly depend on the specific quantum adders that are considered. An analysis of quantum adders for larger numbers of qubits is outside the scope of the present work.

In addition, the compression can be further enhanced with the knowledge of the useful part of the quantum state, namely, the observable or set of observables that are actually relevant for the solution of the problem to deal with. If there is no need of storing and computing the whole quantum state, given as a set of parameters, the autoencoder should be able to efficiently select the important ones. Therefore, apart from the dimensionality analysis, the percentage of the relevant information has to be also taken into account when studying the limits of the compression.
 
 An interesting issue to address in the future is to consider quantum autoencoders based on probabilistic quantum adders~\cite{qa2}, which could also bring novel possibilities, such as larger fidelities at a finite success probability. We finally point out that there exist related possibilities for future paradigms of quantum autoencoders, in the context of quantum circuit optimization, active and reinforcement learning of quantum experiments, as well as supervised learning of Hamiltonians \cite{VenturelliEtAl18,GuerreschiEtAl17,MelnikovEtAl18,InnocentiEtAl18}. The emerging field of quantum adders enhances the paradigm of quantum autoencoders, since the former have already been implemented in current quantum technologies~\cite{qa3,qa3bis,qa3bis2,qa5}. We also remark that an experiment of this proposal in the Rigetti cloud quantum computer has been recently carried out~\cite{YongchengDing}, and another experiment of a classically-optimized, photonic quantum autoencoder has been completed~\cite{NoraTischler}.

\section*{Acknowledgements}
The authors acknowledge support from Spanish MINECO FIS2015-69983-P, Ram\'{o}n y Cajal Grant RYC-2012-11391, UPV/EHU Postdoctoral Grant, Basque Government Postdoctoral Grant POS\_2017\_1\_0022 and IT986-16.

\end{document}